\documentclass[floatfix,reprint,amsmath,amssymb,aps,prb]{revtex4-2}

\usepackage{graphicx}% Include figure files
\usepackage{dcolumn}% Align table columns on decimal point
\usepackage{bm}% bold math
\usepackage{color}
\usepackage[
				colorlinks = true,
	            linkcolor = blue,
	            urlcolor  = blue,
	            citecolor = blue,
	            anchorcolor = blue
	            ]{hyperref}

\newcommand{\E}{\varepsilon}
\newcommand{\h}[1]{\hat{#1}}
\newcommand{\bb}[1]{{\bf #1}}

\newcommand{\hh}{{\bf H}}
\newcommand{\ee}{{\bf E}}

\newcommand{\vv}{{\bf v}}
\newcommand{\rr}{{\bf r}}
\newcommand{\eee}{{\bf e}}

\begin{document}

\title{Non-Hermitian light beams}

\author{Andrey~Novitsky$^{1}$}
\email{andreyvnovitsky@gmail.com}
\author{Dongliang Gao$^{2}$}
\author{Iryna Kaputskaya$^{1}$}
\author{Dawei Li$^{3}$}
\author{Andrey Bogdanov$^{4,5}$}
\author{Lei Gao$^{6,7}$}
\author{Mikhail Rybin$^{5}$}
%\author{Denis~V.~Novitsky$^{7}$}

\affiliation{$^1$ Belarusian State University, Nezavisimosti Avenue
4, 220030 Minsk, Belarus
\\ $^2$ School of Physical Science and Technology \& Jiangsu Key Laboratory of Frontier Material Physics and Devices, Soochow University, Suzhou 215006, China
\\ $^3$ School of Optoelectronic Engineering and
Instrumentation Science, Dalian University of Technology, Dalian
116024, China
\\ $4$ Qingdao Innovation and Development Center, Harbin Engineering University,
Qingdao 266000, Shandong, China
\\ $^5$ ITMO University, Kronverksky Prospekt 49, 197101 St. Petersburg, Russia
\\ $^6$ School of Optical and Electronic Information, Suzhou City University, Suzhou 215104, China
\\ $^7$ Jiangsu Key laboratory of Biophotonics \& Suzhou Key Laboratory of Biophotonics \& SZU-ITU International Joint Meta-center for Advanced Photonics and Electronics, Suzhou City University, Suzhou 215104, China
%\\ $^7$ B. I. Stepanov Institute of Physics, National Academy of Sciences of Belarus, Nezavisimosti Avenue 68, 220072 Minsk, Belarus
}

\begin{abstract}
Non-Hermitian systems provide remarkable features actively studied
in modern photonics. Non-Hermiticity is often related to the
properties of open structures and devices, which involve lossy and
gain channels. Here we reveal that electromagnetic fields themselves
can be designed in a non-Hermitian way proposing an additional
degree of freedom in non-Hermitian physics. Within classical
electromagnetic wave theory we show that the non-Hermitian behaviors
may stem not only from the dispersive properties of the waves, but
also from the peculiarities of the light spatial spectrum and local
structure of the fields. For the nondiffracting light beams we
identify exceptional points detaching symmetric and asymmetric
non-Hermitian phases related to the shape of beam's intensity
profile, the non-Hermiticity being well recognized only for
non-paraxial beams. In the local structure of the electromagnetic
fields and Poynting vector, we determine exceptional lines
differentiating non-Hermitian phases stemming from the behavior of
streamlines. We believe that the concept of non-Hermitian light
beams will enrich our knowledge of the light-matter interaction.
\end{abstract}

%\pacs{78.35.+c, 87.64.Cc}

\maketitle

\section{Introduction}

Propagating light fields (light beams) play a tremendous role in science and everyday life, manifesting themselves in fundamental phenomena like reflection and radiation pressure, as well as in practical applications such as optical manipulation and communication~\cite{Cheng2025,Zhang2025,Omatsu2025,He2025,Forbs2026PiP,Forbs2026PR,Habibovic2024,Zhang2026}.
Beyond the archetypal plane waves, there is a realm of different types of light beams that can be generated using their superposition. Some of these beams as Gaussian, Laguerre-Gaussian, Hermite-Gaussian, and top-hat beams, being realized with the laser radiation in a well-known way, are the solutions to the paraxial Helmholtz equation \cite{Zauderer1986,Pampaloni2004,Kim1999}. More intricate optical fields have been actively studied theoretically and the vast majority of them have been realized in practice. The famous Bessel beam represents a non-diffractive composition of plane waves, whose wavevectors lie on the cone \cite{Durnin1987,Bouchal1998,McGloin2005,Rao2024}. Paraxial Bessel beams can be generated with circular slits and lenses, axicons, metasurfaces, distributed grating arrays, and spatial light modulators (SLMs) representing well established techniques \cite{Akram2019,Kim2026,Zhi2023}. A tunable liquid crystal Fresnel lens was proposed to experimentally generate and characterize vortex singular beams in Ref. \cite{Melnikova2024}. Experimental realization of non-paraxial Bessel beams \cite{Borghi2001,Novitsky2007} is challenging, but they are prospective for dragging particles toward the light source, akin tractor beams \cite{Chen2011,Novitsky2011}. The other diffraction-free partner of the Bessel beam is the Airy light beam that was first introduced in quantum mechanics \cite{Berry1979}. An accelerated Airy light beam travels along a parabolic trajectory \cite{Siviloglou2007} and may need non-paraxial corrections in order to accurately describe the two-dimensional \cite{Minovich2011} and three-dimensional \cite{Li2020} counterparts of the beam. Topological properties of the light fields were identified in a
series of papers \cite{Berry2001,Irvine2008,Arrayas2017,Wang2025} on field knots and links, whose solutions may follow from Bateman's construction or Hopf's fibration. Here, we attempt to expand the zoo of structured light beams with the concept of non-Hermitian beams.

Starting from the seminal paper \cite{Bender1998}, non-Hermitian physics has become a broad area of both theoretical and experimental research spanning quantum mechanics \cite{Bender2007}, photonics
\cite{Feng2017,El-Ganainy2018,Ozdemir2019,Wang2021,Wang2023}, acoustics \cite{Huang2024,Wen2022}, etc. Photonics has become one of the major platforms for exploring non-Hermitian physics. It involves a plethora of promising results such as lasing and anti-lasing \cite{Feng2014,Hodaei2014}, improved sensing \cite{Chen2017,Hodaei2017,Carlo2022,Parto2025,Wiersig2020}, enhanced nonlinear response~\cite{Liu2026NatNano, Ramezanpour2021PRA,Bai2024PRL}, and controlling optical manipulation \cite{Fan2025, Yuan2025} to mention a few.  The difference between Hermitian and non-Hermitian cases consists in the possibility of qualitative change of the behavior of the latter when passing through a distinguished point called the exceptional point (EP) \cite{Zhang2026EPReview,Miri2019,Meng2024,Novitsky2024}. Peculiarity of the EP is caused by simultaneous coalescence of eigenvalues and eigenvectors, the generalized eigenvectors constructed at the EP being non-orthogonal. Non-Hermitian systems may not have a spatial distribution of material parameters as often attributed to PT (parity-time) symmetry. Non-Hermitian operators commonly arise in open photonic systems through radiation, absorption, gain, etc. Beyond this conventional route, recent approaches have demonstrated that non-Hermitian Hamiltonians with prescribed properties can be directly constructed and experimentally implemented, enabling non-Hermiticity to be engineered at the level of the governing operator~\cite{Xiao2026}. Radiative openness alone can generate non-Hermitian spectral singularities. For example, exceptional and diabolic points in subwavelength Mie resonators have been directly related to anapole and superscattering regimes and observed through their scattering spectra~\cite{Zhang2025SciAdv}. 

This research aims to demonstrate that light beams can be constructed with predefined non-Hermitian properties at positions of exceptional points. In fact, the very basic plane wave may qualitatively change its behavior and turn into an evanescent wave. In this case, the wavevector orthogonal to the propagation direction corresponds to the exceptional point. We explicitly determine such an EP by introducing the corresponding non-Hermitian Hamiltonian. In a similar manner, the directions of degeneracy in anisotropic media can be identified. However, the non-Hermiticity in this case is defined by the medium, but not by the light beam itself. Here, we artificially insert a non-Hermitian Hamiltonian into the plane wave decomposition of the light beam. The benefit of such a definition of the beam is that we are able to control the position of the exceptional point. Experimental realization of such a beam may be feasible using the phase engineering through a spatial light modulator or liquid crystal phase masking. We also discuss another way to detect non-Hermiticity in light fields. We consider a local field structure that is described by the linearized system of equations for the electric field, magnetic field, or Poynting vector lines. The matrix derived from the system of linear equations is non-Hermitian, thus, featuring exceptional points and lines and different non-Hermitian phases of state. Previous research on the non-Hermitian light generation is related to the wavefront modulation. In Ref. \cite{Yang2024}, the asymmetric response of EP photonic systems (Pancharatnam-Berry metasurfaces) is employed for creating an arbitrary vectorial wavefront. Non-Hermitian plasmonic metasurfaces with extrinsic chirality result in engineering arbitrarily polarized diffraction EPs that can be used for generating electromagnetic fields with desired properties \cite{Qin2025}. The work \cite{Modak2024} is related to the optical beam response and investigates the phase transition arising in the non-Hermitian beam shift. In this paper, we propose a general theoretical approach for angular-spectrum construction of non-Hermitian light fields that can be further implemented using, e.g., a metasurface or SLM. The approach allows us to systematically investigate the properties of the light beams with imprinted EPs.

The paper is organized as follows. In Section II, following the Introduction, exceptional points due to dispersive properties of plane waves are discussed, a non-Hermitian Hamiltonian is constructed, and exceptional points separating propagating and evanescent waves are identified. In Section III, diffraction-free light beams with embedded non-Hermitian features are investigated. The dependencies on the order of the exceptional point and non-paraxiality are analyzed. Section IV introduces non-Hermitian behaviors by means of the stability matrix for local fields. Derived exceptional line equations for the Poynting vector and the electric field are demonstrated with numerical examples. Section V summarizes the findings of the paper.

\section{Dispersion-induced exceptional points}

Dispersion manifests itself in propagation of plane waves in a
medium. Firstly, we consider a monochromatic electromagnetic wave in
free space that satisfies the wave equation for the electric field
strength $\ee(x,y,z,\omega)$ as follows
\begin{equation}
\frac{\partial^2 \ee}{\partial z^2} + \Delta_\perp \ee + k_0^2 \ee =
0,
\end{equation}
where $\Delta_\perp = \partial^2 /\partial x^2 + \partial^2
/\partial y^2$ is the transverse Laplacian operator, $k_0 =
\omega/c$ is the wavenumber in vacuum, $\omega$ is the angular
frequency, and $c$ is the speed of light.

If we had a light beam propagating along the $z$ axis, the operator
$\Delta_\perp$ would define the transverse structure of the beam.
Propagation of the light beam along the $z$ axis can be
characterized with the propagation constant (longitudinal
wavenumber) $\beta$. This translation symmetry implies the
$\exp(i\beta z)$-dependence of the electric field. Hence, the wave
equation for the electric field $\ee(x,y) = f(x,y) {\bf e}$ can be
rewritten as
\begin{equation}
\Delta_\perp f = -\eta^2 f, \label{EqFor_f}
\end{equation}
where $\eta^2 = k_0^2 - \beta^2$ and ${\bf e}$ is a polarization
vector. Due to the plane electromagnetic wave transversality
condition ${\bf k} \cdot \ee = 0$ the polarization vector can be
presented as ${\bf e} = {\bf A}\times {\bf k}/k_0$, where ${\bf A}$
is a vector amplitude that can generally depend on the wavevector
${\bf k}$. Hence, the electric field of the plane wave reads as
\begin{equation}
\ee_{\bf k}(x,y) = [{\bf A}({\bf k})\times {\bf k}/k_0] f(x,y).
\label{ElFieldAxkf}
\end{equation}

The left-hand side of Eq. (\ref{EqFor_f}) can be factorized. Then
the equation reads as
\begin{equation}
\left( \frac{\partial}{\partial x} + i \frac{\partial}{\partial y}
\right) \left( \frac{\partial}{\partial x} - i
\frac{\partial}{\partial y} \right) f = -\eta^2 f.
\end{equation}

This factorization yields the system of the first-order equations
\begin{eqnarray}
&& \left( \frac{\partial}{\partial x} - i \frac{\partial}{\partial
y}
\right) f = g, \nonumber \\
&& \left( \frac{\partial}{\partial x} + i \frac{\partial}{\partial
y} \right) g = -\eta^2 f
\end{eqnarray}
or equivalent Schr\"{o}dinger-like equation
\begin{equation}
i \frac{\partial \psi}{\partial x} = \hat H \psi,
\end{equation}
where the wave function and effective Hamiltonian are equal to
respectively
\begin{eqnarray}
\psi = \left( \begin{array}{c} f \\ g \end{array} \right), \quad
\hat H = \left( \begin{array}{cc} -\frac{\partial}{\partial y} & i \\
-i\eta^2 & \frac{\partial}{\partial y}
\end{array} \right). \label{Hamilt1D}
\end{eqnarray}

For the harmonic dependence $\exp(i k_y y)$ of functions $f$ and $g$
we arrive at the Hamiltonian
\begin{eqnarray}
\hat H = \left( \begin{array}{cc} -i k_y & i \\
-i \eta^2 & i k_y \end{array} \right).
\end{eqnarray}
In general, the wavenumber $k_y$ is complex showing either decay or
growth of the fields in the $y$ direction, but the Hamiltonian is
non-Hermitian for real-valued $k_y$, too. Furthermore, we inspect
the eigenvalues and eigenvectors of such a Hamiltonian. The
eigenvalues are equal to $\lambda_{1,2} = \pm \sqrt{\eta^2 - k_y^2}$
and have the meaning of the wavenumber $k_x$. The eigenvectors have
the following form
\begin{eqnarray}
\psi_{1,2} \sim \left( \begin{array}{c} i  \\
i k_y \pm \sqrt{\eta^2 - k_y^2} \end{array} \right).
\end{eqnarray}

\begin{figure}[tb]
  \centering
  \includegraphics[width=0.9\linewidth]{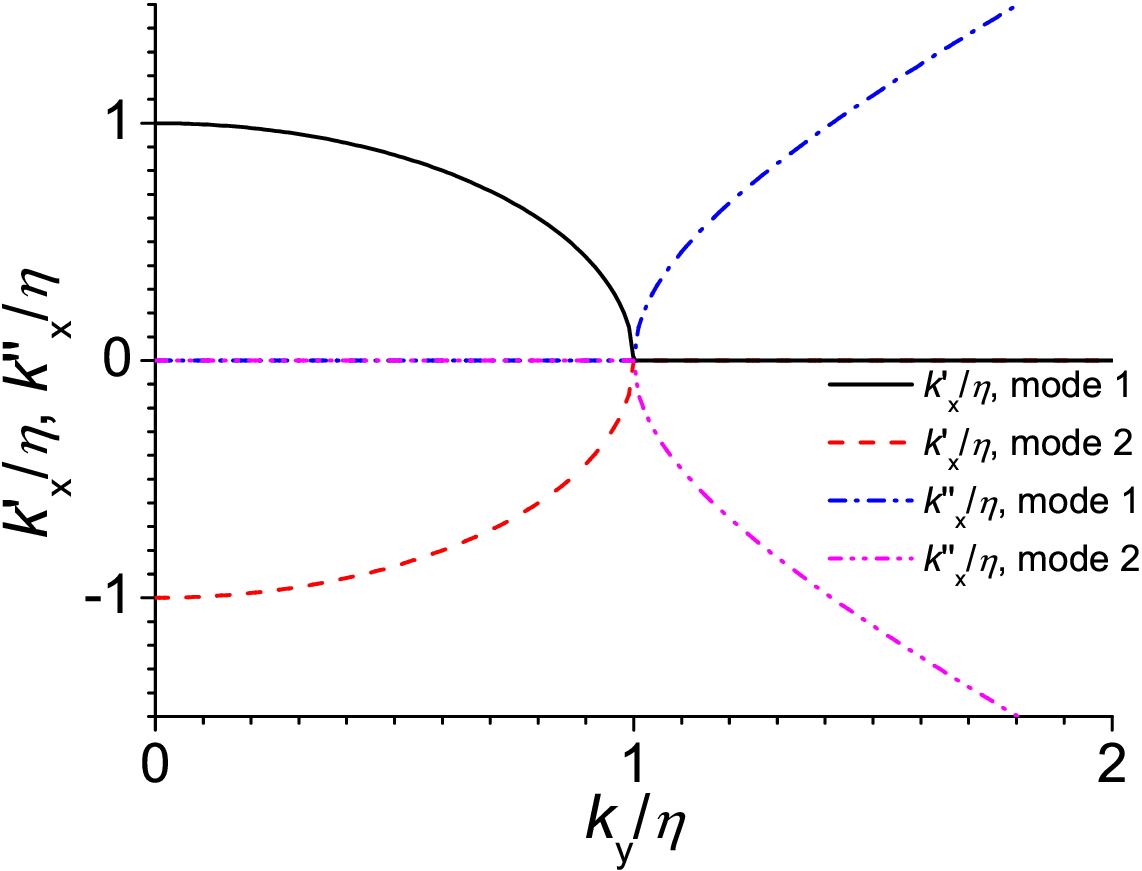}
  \caption{Real and imaginary parts of the eigenvalues for modes $k_x = + \sqrt{\eta^2 - k_y^2}$ and $k_x = - \sqrt{\eta^2 - k_y^2}$.
  Only the positive branch is displayed with the EP at $k_y = +\eta$.
  This EP corresponds to the auxiliary first-order evolution operator \ref{Hamilt1D} obtained by factorizing the Helmholtz equation.}
  \label{fig1}
\end{figure}

Eigenvalues and eigenvectors coalesce, if $|k_y| = \eta$. This is an
exceptional point of the plane wave. Such an exceptional point is
quite trivial for real-valued wavenumbers $k_y$, because it
separates propagating ($|k_y| < \eta$, $k_x$ is real) and evanescent
($|k_y|>\eta$, $k_x$ is imaginary) modes. Since $x$ and $y$
derivatives can be exchanged in the Schrodinger equation, the EP
also exists at $|k_x| = \eta$.
%Combining these cases, we may
%conclude that EPs lie on the circle $k_x^2 + k_y^2 = \eta^2$, which
%is the dispersion equation for waves in free space.
In Ref. \cite{Wu2026} such an EP has been exploited when discussing
optical pulling forces in propagating and evanescent regimes. Figure
\ref{fig1} clearly shows the emergence of the exceptional point at
$k_y = \eta$, when the eigenvalues (wavenumbers $k_x$) of two modes
coalesce. It is clear that the EP demarcates the regions of
propagating and evanescent waves in the case of real values.

If $k_y = k_y' + i k_y''$ is complex, then the EP can appear either
due to appropriate value of complex $\beta$ or due to the complex
medium permittivity $\varepsilon$. Since $\varepsilon=1$, we
consider the case of complex $\beta = \beta' + i \beta''$. The
condition for emergence of the EP reads as
\begin{eqnarray}
k_0^2 -(\beta' + i \beta'')^2 - (k_y' + i k_y'')^2 = 0
\end{eqnarray}
or
\begin{eqnarray}
k_0^2  + \beta^{\prime\prime 2} + k_y^{\prime\prime 2} =
\beta^{\prime 2} + k_y^{\prime 2}, \qquad \beta' \beta'' + k_y'
k_y'' = 0.
\end{eqnarray}

Thus, the real and imaginary parts of the wavenumber $k_y$ defining
the EP are equal to
\begin{eqnarray}
k_y^\prime = \pm \sqrt{\frac{ b + \sqrt{b^2 + 4 \beta'
\beta''}}{2}}, \quad
k_y^{\prime\prime} = - \frac{\beta'\beta''}{k_y^\prime},
\end{eqnarray}
where $b = k_0^2 + \beta^{\prime\prime 2} - \beta^{\prime 2}$.

%In Fig. \ref{fig1} we show the wavenumbers $k_y^{\prime} > 0$ and
%$k_y^{\prime\prime}$ as a function of $\beta'>0$ and $\beta''>0$. In
%case $\beta''=0$, the real part $k_y^{\prime}$ is a segment of a
%circle for $\beta'<1$ and line on the abscissa axis for $\beta'>1$.
%The compensation of losses along the $z$-axis needs the gain
%$k''_y<0$ along the $y$-axis. Such a balance results in the
%appearance of the EP. For greater $\beta''$ the anticrossing of
%lines of real and imaginary parts of the wavenumber $k_y$ is
%observed in Fig. \ref{fig1}.

%\begin{figure}[tb]
%  \centering
%  \includegraphics[width=0.9\linewidth]{Fig1.eps}
%  \caption{Real and imaginary parts of the wavenumber $k_y$. }
%  \label{fig1}
%\end{figure}

Further we will look at the behavior of the polarization vector
${\bf e} = {\bf A}\times {\bf k}/k_0$ when crossing an exceptional
point. Taking ${\bf A} = A(k_y) {\bf e}_z$ as an example, we get
${\bf e} = A(k_y) (k_x {\bf e}_y - k_y {\bf e}_x)/k_0$, where ${\bf
e}_{x}$, ${\bf e}_{y}$, and ${\bf e}_{z}$ are the basis vectors of
Cartesian coordinates. In the propagating mode ($k_y < \eta$), the
plane wave is linearly polarized. In the evanescent mode ($k_y
> \eta$), the wave has imaginary value of $k_x = i \sqrt{k_y^2 -
\eta^2}$ and, therefore, its polarization is elliptical: ${\bf e} =
A(k) \left( i \sqrt{k_y^2 - \eta^2} {\bf e}_y - k_y {\bf e}_x
\right)/k_0$. At the exceptional point $|k_y| = \eta$, the
polarization is linear and oriented along the $y$ axis.

If the wavenumber $k_z$ is not fixed with the propagation constant
$\beta$, the problem is three-dimensional and the Hamiltonian
entering the Schr\"{o}dinger equation
\begin{equation}
i \frac{\partial \psi}{\partial z} = \hat H \psi \label{SchrodEq_z}
\end{equation}
equals
\begin{eqnarray}
\hat H = \left( \begin{array}{cc} -\sqrt{\Delta_\perp} & i \\
-i k_0^2 & \sqrt{\Delta_\perp}
\end{array} \right),
\end{eqnarray}
where the square root of the differential operator is formally
introduced.

Taking the harmonic dependence $\exp(ik_x x + i k_y y)$ of functions
$f$ and $g$, we get
\begin{eqnarray}
\hat H = \left( \begin{array}{cc} -i \sqrt{k_x^2 + k_y^2} & i \\
-i k_0^2 & i \sqrt{k_x^2 + k_y^2}
\end{array} \right).
\end{eqnarray}
The eigenvalues $\lambda_{1,2} = \pm \sqrt{k_0^2 - k_x^2 - k_y^2}$
are the wavenumbers $k_z$. They separate propagating and evanescent
solutions. On the plane ($k_x$, $k_y$) we have an exceptional line
formed by EPs on the circle $k_x^2 + k_y^2 = k_0^2$. Thus, the EPs
are introduced as a consequence of the dispersive properties of the
plane wave, which are simple in free space, but become more complex
in a material medium.

To determine exceptional points for electromagnetic waves in
bi-anisotropic media, we start with the Schr\"{o}dinger-like
equation for tangential electric $\ee_t = I \ee$ and magnetic $\hh_t
= I \hh$ fields (a couple of two-component vectors), where $I =
I_3-{\bf e}_z\otimes{\bf e}_z$ is the projecting operator onto the
plane orthogonal to the $z$-axis and $I_3$ is the three-dimensional
unit matrix. Then the wavefunction in (\ref{SchrodEq_z}) is a
4-component vector $\psi = (\hh_t, {\bf e}_z \times \ee)^T$. The
effective Hamiltonian following from the Maxwell equations equals
\begin{equation}
\h H = - k_0 \left( \begin{array}{cc} A & B \\ C & D \end{array}
\right).
\end{equation}
The block matrices A, B, C, and D are provided in Appendix
\ref{AppA}. Eigenvalues $\lambda$ of the Hamiltonian $\h H$
represent the wavenumbers $k_z$. They take the same values at the
boundary between the propagating and evanescent waves giving rise to
exceptional points. In (bi)anisotropic media, both ordinary and
extraordinary waves exist. Intersection of the dispersion surfaces
of both types of waves does not result in the appearance of the EPs,
because these waves have different polarizations (different
eigenvectors). Instead, one observes diabolic points. Nevertheless,
in biaxial anisotropic crystals and bianisotropic media the EPs can
emerge. They correspond to particular directions, in which linear,
quadratic, and cubic coordinate dependencies of the fields are
realized \cite{Borzdov1996}. Exceptional points are also analyzed in
photonic crystals as, e.g., in Ref. \cite{Rybin2016}.

%The wavenumbers can be presented in the other parametric form. When
%we transfer from Cartesian to polar coordinates, the wavenumbers are
%$k_x = K \cos\theta$ and $k_y = K \sin\theta$.

\section{EP imprinted in the spatial structure of the light beam}

For a separable quantum state in quantum mechanics, the wavefunction
is presented as a product of the coordinate-dependent wave function
$\varphi(\rr)$ satisfying the Schr\"{o}dinger equation and
spin-dependent wave function $\chi({\bf s})$ as follows
$\psi(\rr,{\bf s}) = \varphi(\rr) \chi({\bf s})$. Similarly, for
electromagnetic plane waves, the coordinate-dependent function reads
as $\varphi(\rr)=\exp(i{\bf k} \cdot \rr)$, while the
polarization(spin)-dependent function is $\chi={\bf A}({\bf
k})\times {\bf k}/k_0$. The wavefunction of the plane
electromagnetic wave is $\psi = [{\bf A}({\bf k})\times {\bf k}/k_0]
\exp(i{\bf k} \cdot \rr)$.

As we have seen before, exceptional points of the
coordinate-dependent function $\varphi(\rr)$ are related to the
dispersive properties of electromagnetic waves separating
propagating and evanescent solutions. Polarization-dependent
function $\chi$ may also generate EPs embedded into the vector ${\bf
A}$. Indeed, this amplitude vector can be presented as ${\bf A} =
F(\hat H){\bf A}_0$, where $F$ is an arbitrary function of a
non-Hermitian Hamiltonian $\hat H$ and ${\bf A}_0$ is a vector. Both
$\hat H$ and ${\bf A}_0$ can be functions of the wavevector ${\bf
k}$. The Hamiltonian is a $3\times 3$ matrix with eigenvalues
$\lambda_1$, $\lambda_2$, and $\lambda_3$. At any point except the
EP, the Hamiltonian can be expanded into spectral series $\hat H =
\sum_{j=1}^3 \lambda_j \rho_j$, where $\rho_j={\bf v}_j \otimes {\bf
u}_j$ is a tensor product of left and right eigenvectors of the
Hamiltonian with the biorthogonal normalization condition ${\bf v}_j
\cdot {\bf u}_j = \delta_{ij}$, $\hat H {\bf v}_j = \lambda_j {\bf
v}_j$ and ${\bf u}_j \hat H = \lambda_j {\bf u}_j$. In the case of
the second-order EP, two eigenvalues and eigenvectors coalesce. One
of the examples of such a Hamiltonian is
\begin{eqnarray}
\hat H = k_0^{-1} \left( \begin{array}{ccc} k_y - i \gamma_1 & \kappa & 0\\
\kappa & k_y - i\gamma_2 & 0 \\ 0 & 0 & k_0
\end{array} \right). \label{H2}
\end{eqnarray}
Here we introduce the dependence in the effective Hamiltonian on the
wavenumber $k_y$ and three parameters $\kappa$, $\gamma_1$, and
$\gamma_2$. The structure of the Hamiltonian (\ref{H2}) is similar
to that adopted for description of the system of two coupled
resonators \cite{Doppler2016} [it is a two-dimensional matrix in the
left upper corner of the matrix (\ref{H2})]. The eigenvalues of
$\hat H$ are $\lambda_{1,2} = (k_y - i (\gamma_1 + \gamma_2)/2 \pm
\sqrt{\kappa^2 -(\Delta\gamma/2)^2})/k_0$ and $\lambda_3 = 1$ with
$\Delta\gamma = |\gamma_1-\gamma_2|$. Two eigenvalues and two
eigenvectors coincide at the exceptional point, when $\kappa =
\Delta\gamma/2$, and equal $\lambda_{EP} = k_y/k_0 - i (\gamma_1 +
\gamma_2)/2k_0$ and ${\bf
v}_{EP}=(1,i(\gamma_1-\gamma_2)/\Delta\gamma,0)^T$, respectively.
The amplitudes in both non-Hermitian phases separated by the
exceptional point can be written using the matrix spectral
decomposition as ${\bf A} = \sum_{j=1}^3 F(\lambda_j) \rho_j{\bf
A}_0$.

At the exceptional point, the Hamiltonian is of the form $\hat
H_{EP} = \lambda_{EP} I_2 + \lambda_3 \rho_3 + \hat N$, where $I_2 =
I_3 - \rho_3$, $I_3$ is the three-dimensional unit matrix, and $\hat
N$ is the nilpotent matrix ($\hat N^2 = 0$). Then we arrive at the
amplitude ${\bf A}_{EP} = [F(\lambda_{EP}) I_2 + F(\lambda_3) \rho_3
+ F'(\lambda_{EP}) \hat N] {\bf A}_0$, where $F'$ is the derivative
of the function $F$.

\begin{figure}[tb]
  \centering
  \includegraphics[width=\linewidth]{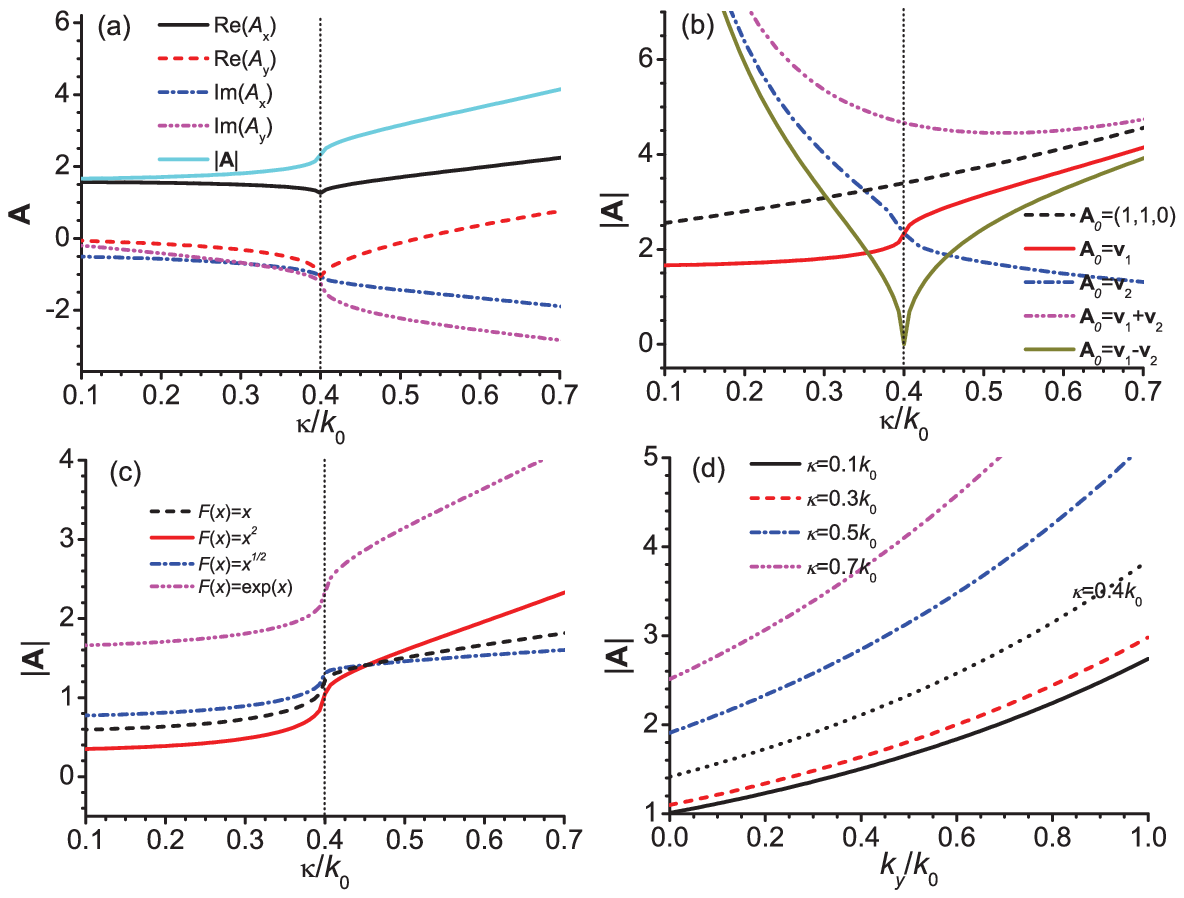}
  \caption{(a) Components of the vector amplitude ${\bf A}$ and its
  norm $|{\bf A}|$ depending on the parameter $\kappa$.
  Norm of the amplitude $|{\bf A}|$ for different (b) vector parameters ${\bf
  A}_0$ and (c) functions $F(x)$. Norm $|{\bf A}|$ as a function of $k_y$
  for different $\kappa$.  Default parameters: $\gamma_1 = 0.3 k_0$,
  $\gamma_2 = 1.1 k_0$, $k_z = 0.5 k_0$, $F(x) = \exp(x)$, and ${\bf A}_0 = {\bf
  v}_1$.}
  \label{fig2}
\end{figure}

\begin{figure*}[tb]
  \centering
  \includegraphics[width=\linewidth]{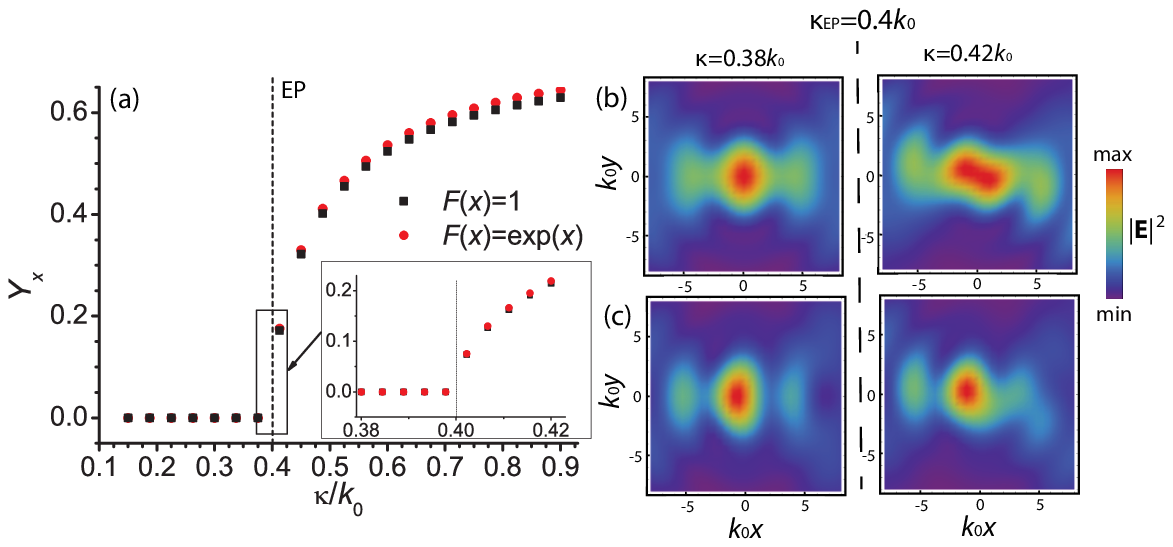}
  \caption{ (a) Asymmetry parameter $Y_x$ versus parameter $\kappa$ for two functions $F(x)$.
  Zoomed-in region near the EP is shown in the inset.
  Light beam's intensity for $\kappa < \kappa_{EP}$ and $\kappa > \kappa_{EP}$ and functions (b) $F(x) = 1$ and (c) $F(x) = \exp(x)$.
  Calculation parameters: $\gamma_1 = 0.3 k_0$, $\gamma_2 = 1.1 k_0$,
  $k_z = 0.5 k_0$, and ${\bf A}_0 = {\bf v}_1$.
  Vertical dashed line corresponds to the position of the exceptional point. Intensities in the panels are normalized independently.}
  \label{fig3}
\end{figure*}

Behaviors of the vector amplitude ${\bf A}$ are analyzed in Fig.
\ref{fig2}. In Fig. \ref{fig2}(a), we show the real and imaginary
components of the vector amplitude ${\bf A}$, when ${\bf A}_0 = {\bf
v}_1$ (${\bf A}$ is an eigenvector). For the chosen parameters, the
exceptional point corresponds to $\kappa_{EP} = 0.4 k_0$ being
marked with the vertical dotted line. We can clearly see that the
curves experience the change of their shape at the EP. The real
parts have cusps, while imaginary parts bend. A peculiar behavior
--- an inflection point --- is observed also in the norm of the
vector amplitude $|{\bf A}|$. In Fig. \ref{fig2}(b) we study the
effect of the vector ${\bf A}_0$. In the case of just a constant
vector one does not see any peculiarities in the line of $|{\bf A}|$
(dashed curve). However, a superposition of eigenvectors ${\bf
v}_{1,2}$ may result in a feature, which is clearly observed in all
cases except ${\bf A}_0 = {\bf v}_1 + {\bf v}_2$. The curves for
${\bf A}_0 = {\bf v}_1$ and ${\bf A}_0 = {\bf v}_2$ cross each other
at the EP, because the corresponding amplitudes ${\bf A} =
F(\lambda_1) {\bf v}_1$ and ${\bf A} = F(\lambda_2) {\bf v}_2$
coincide there. This is also the reason for the cusp at ${\bf A}_0 =
{\bf v}_1 - {\bf v}_2$. Then we investigate the influence of the
function $F(x)$ on the line shape near the EP. In Fig.
\ref{fig2}(c), we consider four functions and see that all of them
keep inflection point. For $\kappa < \kappa_{EP}$, all curves behave
similarly independent of $F(x)$. The function manifests itself in
the range $\kappa > \kappa_{EP}$: The weakest and strongest
dependencies correspond to the square-root and exponential
functions, respectively. Different behaviors in non-Hermitian phases
are also seen in the $k_y$-dependence in Fig. \ref{fig2}(d). For
$\kappa < \kappa_{EP}$, the lines rise up slowly accelerating near
$\kappa_{EP} = 0.4 k_0$. After exceeding $\kappa_{EP}$ the distance
between curves increases. Further we will look at the impact of the
identified non-Hermiticity on the propagating light field (light
beam).

\begin{figure*}[tb]
  \centering
  \includegraphics[width=\linewidth]{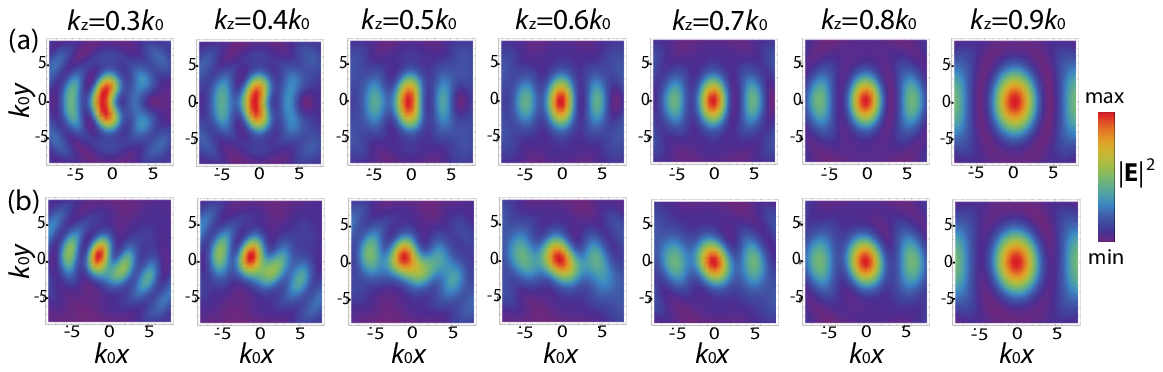}
  \caption{ Light beam's intensity for different $k_z$ and
  (a) $\kappa = 0.3 k_0$ (below EP) and (b) $\kappa = 0.5 k_0$ (above EP).
   Parameters: $\gamma_1 = 0.3 k_0$, $\gamma_2 = 1.1 k_0$,
  $F(x) = \exp(x)$, and ${\bf A}_0 = {\bf v}_1$. Intensities in the panels are normalized independently.}
  \label{fig4}
\end{figure*}

Light beam can be composed of elementary plane waves described
above. Since non-Hermitian spatial spectrum is imprinted into the
amplitudes ${\bf A}$, the light beam might have different behaviors
in non-Hermitian phases separated by the exceptional point. Electric
field of the light beam consisting of propagating plane waves can be
written as follows
\begin{equation}
\ee(\rr) = e^{ik_z z} \sum_{\sigma=\pm 1} \int_{-\eta}^\eta [{\bf
A}(k_y)\times {\bf k}_\sigma/k_0] e^{i \sigma k_x x + i k_y y} dk_y,
\label{beamField}
\end{equation}
where $k_z$ is fixed and $k_x = \sqrt{\eta^2 - k_y^2}$, and
$\textbf{k}_\sigma = (\sigma k_x,k_y,k_z)^T$. For such a
nondiffracting field, the light intensity $|\ee(x,y)|^2$ is a
function of transverse coordinates.

\begin{figure*}[tb]
  \centering
  \includegraphics[width=\linewidth]{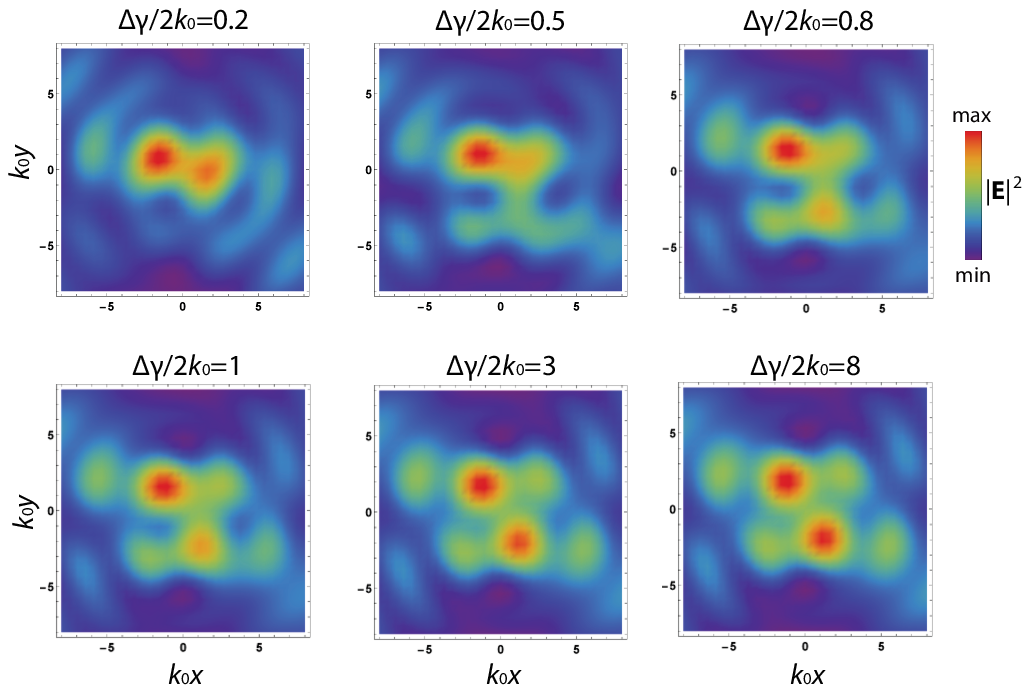}
  \caption{ Light beam's intensity depending on the difference $\Delta\gamma$
  for the Hamiltonian $\hat H'$.
  Parameters: $\gamma_1 = 0.3 k_0$,
  $k_z = 0.5 k_0$, $\kappa = 0.3 k_0$, $F(x) = 1$, and ${\bf A}_0 = {\bf v}_1$. Intensities in the panels are normalized independently.}
  \label{fig5}
\end{figure*}

\begin{figure}[tb]
  \centering
  \includegraphics[width=\linewidth]{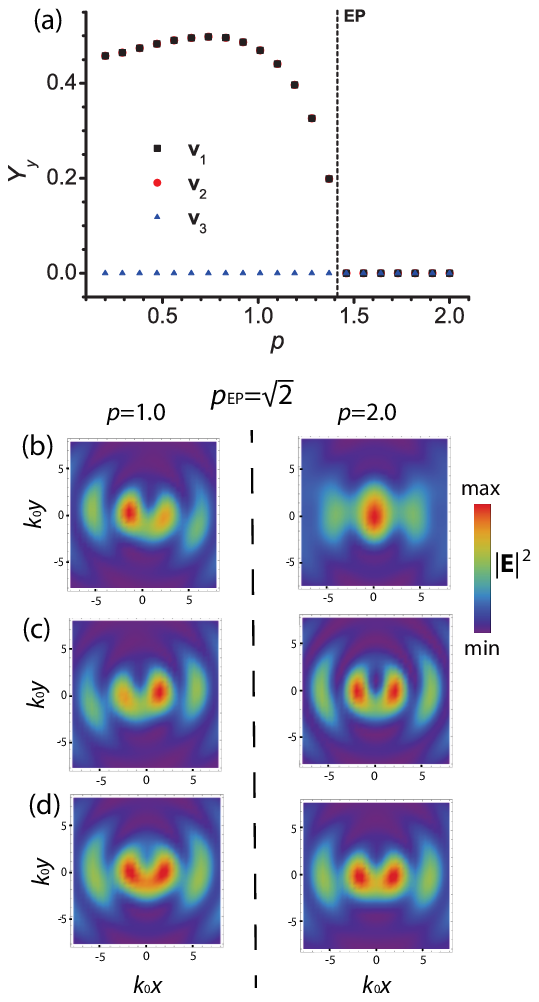}
  \caption{ (a) Asymmetry parameter $Y_y$ versus parameter $p =
  g/\kappa$ in the case of the Hamiltonian $\hat H_3$
  for different vectors ${\bf A}_0$. Light beam's intensity
  depending on the ratio $p = g/\kappa$ for (b) ${\bf A}_0 = {\bf v}_1$,
  (c) ${\bf A}_0 = {\bf v}_2$, and (d) ${\bf A}_0 = {\bf v}_3$. Parameters: $\kappa = 0.5 k_0$,
  $k_z = 0.5 k_0$, and $F(x) = 1$. Intensities in the panels are normalized independently.}
  \label{fig6}
\end{figure}

In Figs. \ref{fig3}-\ref{fig7}, the intensities in the panels are
normalized independently to demonstrate the spatial distributions,
but not the comparison of intensities in different panels. In Fig.
\ref{fig3}, one can see what happens with the intensity profile
$|\ee|^2$ when changing parameter $\kappa$ at constant $\gamma_1$
and $\gamma_2$. In this case, there are two non-Hermitian phases
merging at the exceptional point $\kappa_{EP} = \Delta\gamma/2$. The
phase at $\kappa<\kappa_{EP}$ can be called symmetric, because the
profile $I(x,y)$ is mirror-symmetric with respect to the axis $x$
for functions $F(x)$, viz., $I(x,y) = I(x,-y)$. The phase at
$\kappa>\kappa_{EP}$ is asymmetric as it clearly follows from the
observations. To quantitatively describe the phase transition at the
EP, we propose the asymmetry parameter:
\begin{equation}
Y_x = \frac{\int_{-L}^L \int_{-L}^L
|I(x,y)-I(x,-y)|dxdy}{\int_{-L}^L \int_{-L}^L I(x,y)dxdy},
\end{equation}
where $2L\times 2L$ is the size of the considered transverse area
[$L = 8/k_0$ in Figs. \ref{fig3}(b) and (c)].

Figure \ref{fig3}(a) shows the change of the asymmetry parameter
$Y_x$ depending on the value of $\kappa$, the results for two
functions $F(x)$ being quite close one to another. Figure
\ref{fig3}(a) with the inset of zoomed-in region near the EP
convinces one in the abrupt change of the behavior when passing the
phase transition point (exceptional point). The asymmetry $Y_x$
equals zero (profile is symmetric) for $\kappa < \kappa_{EP}$ and
monotonously increases (profile is asymmetric) for $\kappa >
\kappa_{EP}$. Symmetric profile has a main high-intensity lobe and a
couple of side lobes. The main lobe changes from elongated shape at
small $\kappa$ to the more round one at $\kappa$ near the EP (Fig.
1S in Ref. \cite{Supplementary} contains more intensity profiles for
different $\kappa$). In vicinity of the EP the lobes blur and
overlap one another. In the asymmetric phase, the main lobe goes to
the two-peak shape, the left main and side lobes being higher than
the right ones. For larger $\kappa$, the two maxima are more clearly
distinguishable. The described evolution of the intensity profiles
is well seen in Fig. \ref{fig3}(b), that is, for $F(x) = 1$, when
the vector amplitude represents the eigenvector of the non-Hermitian
Hamilton ${\bf A} = {\bf v}_1$. In this case, the profile is
mirror-symmetric with respect to both $x$ and $y$ axes. The other
functions $F(x)$ modify this simple behavior: The side lobes are
less pronounced, while the main lobe may deform even in symmetric
phase for some functions $F(x)$ keeping the mirror-symmetry only
with respect to the $x$ axis as in Fig. \ref{fig3}(c). Two peaks in
the main lobe in the asymmetric phase are not equivalent anymore.
The left peak is brighter, while the right peak may not even be
observed due to the modulation of the vector amplitude ${\bf A} =
F(\lambda_1) {\bf v}_1$.

We have considered quite a small value of $k_z = 0.5 k_0$ in Fig.
\ref{fig3}. In other words, we have dealt with a nonparaxial light
beam. Now we would like to answer the question, whether paraxial
beams can show non-Hermitian features. To this end, we calculate the
light beam's intensity in symmetric and asymmetric phases [Figs.
\ref{fig4}(a) and (b), respectively] for different values of $k_z$.
We observe that strongly nonparaxial fields demonstrate better
distinction between symmetric and asymmetric phases. The asymmetry
is clearly visible to the eye and the lobes are slightly rotated.
However, when the longitudinal wavenumber $k_z$ increases, the
asymmetry in Fig. \ref{fig4}(b) fades away and is almost lost at
$k_z = 0.9 k_0$. Thus, the non-Hermitian properties are well
noticeable only for nonparaxial light beams.

The choice of the vector amplitude ${\bf A}_0$ changes the intensity
both in symmetric and asymmetric phases. In particular, Fig. 2S in
Ref. \cite{Supplementary} features the mirror-reflected
distributions for ${\bf A}_0={\bf v}_1$ and ${\bf A}_0={\bf v}_2$.

Another non-Hermitian Hamiltonian can be obtained from Eq.
(\ref{H2}), if $k_y$ and $\kappa$ are swapped:
\begin{eqnarray}
\hat H' = k_0^{-1} \left( \begin{array}{ccc} \kappa - i \gamma_1 & k_y & 0\\
k_y & \kappa - i\gamma_2 & 0 \\ 0 & 0 & k_0
\end{array} \right). \label{H2a}
\end{eqnarray}
The eigenvalues and eigenvectors stay the same as for the
Hamiltonian (\ref{H2}) up to substitution $k_y \leftrightarrow
\kappa$. However, the exceptional point is located in the spatial
spectrum now being equal to $k_y^{EP} = \pm \Delta\gamma/2$. Since
the integration range in the intensity (\ref{beamField}) spans from
$-\eta$ to $\eta$, the EP can be either in the integration domain or
not. As demonstrated in Fig. \ref{fig5}, this influences the light
beam's profile. For the sake of illustration, we consider a simple
case of $F(x) = 1$. The panels show the intensity distributions
depending on position of the EP defined by $\Delta\gamma$ for $\eta
\approx 0.866 k_0$. The upper row in Fig. \ref{fig5} corresponds to
$-\eta < k_y^{EP}< \eta$. There are no EPs in the integration range
for lower row of panels. When $k_y^{EP}$ is small, we have a wide
range of the first non-Hermitian phase and a narrow range
$-|k_y^{EP}| < k_y < |k_y^{EP}|$ of the second non-Hermitian phase.
In other words, the beam is approximately in the first phase, which
is characterized by two spots of close intensity (main lobe) as
demonstrated in the left upper panel of Fig. \ref{fig5}. When
$\Delta\gamma$ rises, the intensity of the right spot degrades. Then
the intensity ``leaks'' downwards forming another spot of low
intensity at near $k_y^{EP} = \pm\eta$, where the beam is entirely
in the second non-Hermitian phase. Further displacement of the EP
towards greater values of $\Delta\gamma$ as in the lower row in Fig.
\ref{fig5} brings us to the pure second phase, which intensity
profile shows a couple of two equal-intensity spots as in the first
phase, but their positions are different. One more distinction from
the intensity profile in the first phase is the emergence of four
side spots (side lobes for each hot spot).

Exceptional points of the 3rd order can be introduced using a
three-dimensional non-Hermitian Hamiltonian, e.g., as in Ref.
\cite{Hodaei2017}
\begin{eqnarray}
\hat H_3 = k_0^{-1} \left( \begin{array}{ccc} k_y + i g & \kappa & 0\\
\kappa & k_y & \kappa \\ 0 & \kappa & k_y - i g
\end{array} \right). \label{H3}
\end{eqnarray}
Such a Hamiltonian has eigenvalues $\lambda_{1,2} = (k_y \pm
\sqrt{2\kappa^2 - g^2})/k_0$, $\lambda_3 = k_y/k_0$ and eigenvectors
${\bf v}_j = \left( \frac{\kappa}{\lambda_j - k_y - i g}, 1,
\frac{\kappa}{\lambda_j - k_y + i g} \right)^T$, where $j=1,2,3$.
The EP of the 3rd order appears at $g = \sqrt{2}\kappa$: Three
eigenvalues and three eigenvectors coalesce and become respectively
$\lambda_{EP} = k_y/k_0$ and ${\bf v}_{EP} = \left( i \kappa/g, 1,
-i \kappa/g \right)^T$. We can again exploit the vector amplitude in
the form ${\bf A} = F(\hat H_3){\bf A}_0$, since the Hamiltonian is
still three-dimensional.

In Fig. \ref{fig6}, we change the ratio of parameters $p = g/\kappa$
to observe two non-Hermitian phases demarcated by the exceptional
point $p_{EP} = \sqrt{2}$. The parameter
\begin{equation}
Y_y = \frac{\int_{-L}^L \int_{-L}^L
|I(x,y)-I(-x,y)|dxdy}{\int_{-L}^L \int_{-L}^L I(x,y)dxdy},
\end{equation}
exhibiting the mirror-asymmetry with respect to the axis $y$ can be
exploited in this case. In Fig. \ref{fig6}(a), the asymmetry is
demonstrated for three vectors ${\bf A}_0$ taking the values of
eigenvectors ${\bf v}_{1,2,3}$ of the Hamiltonian $\hat H_3$. The
asymmetry existing for lower values of $p$ is clearly observed for
${\bf A}_0 = {\bf v}_{1,2}$. However, it is not realized for ${\bf
A}_0 = {\bf v}_{3}$. This can be traced in the evolution of the
intensity profiles shown in Fig. \ref{fig6}(b)-(d) and Fig. 3S in
\cite{Supplementary}. At $p = 1$, we have three different intensity
profiles for different ${\bf A}_0$. Two of them are mirror-reflected
with respect to the axis $y$, while the third one is symmetric.
Approaching the EP ($p = 1.4$) the asymmetry disappears and all
intensity profiles become similar. This asymmetric phase turns into
the symmetric one for $p > p_{EP}$. The most distinct behavior in
symmetric phase is realized in Fig. \ref{fig6}(b): The spots of the
main lobe come closer and further form a single hot spot at the
center. For other ${\bf A}_0$ the spots of the main lobe rather
split more, but do not come together.

In order to introduce an $n$th order EP we need a non-Hermitian
Hamiltonian $\hat H_n$ as at least an $n\times n$ matrix. The
dimensionality has to be reduced to three dimensions to write a
vector ${\bf A}$. This can be done as follows. Vector amplitude can
be presented as ${\bf A} = \hat B F(\hat H_n) {\bf A}_0$, where
$\hat B$ is a rectangular $3\times n$ matrix and ${\bf A}_0$ is an
$n$-dimensional vector (for example, an eigenvector of the
Hamiltonian).

\section{Non-Hermitian near-fields}

In the previous section, we have considered non-Hermitian light
beams propagating over long distances, the non-Hermiticity of which
was hidden in the spatial spectrum. Here we will uncover
non-Hermitian properties of near fields caused by the local field
distribution. The idea we employ follows the results delivered in
Ref. \cite{Novitsky2009}. A fictitious point is supposed to move
along the field line related to a field ${\bf F}(\rr)$ (electric
field, magnetic field, or Poynting vector). The field can be
linearized in the vicinity of a point in question $\rr_0$ using the
series expansion, so that the equation of motion of the fictitious
point reads as
\begin{equation}
\frac{d\rr}{d\tau} = {\bf F}(\rr) \approx {\bf F}(\rr_0) + \hat G_F
(\rr - \rr_0),
\end{equation}
where $\tau$ is a ``time'' parameter and $\hat G_F$ is a stability
matrix. Presenting the radius-vector as $\rr = {\bf \rho} + \rr_0$,
we rewrite the equation of motion near the point $\rr_0$ as follows
\begin{equation}
\frac{d{\bf \rho}}{d\tau} = {\bf F}(\rr_0) + \hat G_F {\bf \rho}.
\label{G_F}
\end{equation}
Stability matrix is generally a three-dimensional matrix. It
consists of the first partial derivatives as
\begin{equation}
\hat G_F = [\nabla_0 \otimes {\bf F}(\rr_0)]^T,
\end{equation}
where $\nabla_0 = \partial/\partial\rr_0$. In index notation, the
stability matrix equals $(\hat G_F)_{ij} = \partial F_i/\partial
x_{0j}$. Thus, the stability matrix unveils the near-field behavior
of the fields.

Equation (\ref{G_F}) resembles the Schr\"{o}dinger equation with the
stability matrix instead of the Hamiltonian. Electromagnetic field
defines the stability matrix; therefore, the properties of the
stability matrix establish the behaviors of the fields, for example,
the availability of the singular points. When the stability matrix
is non-Hermitian, one can get a non-Hermitian dynamics of the
fictitious point and determine the exceptional points of the near
fields. The matrices for the complex-valued electromagnetic fields
are complex, however, the stability matrix for the Poynting vector
${\bf S} = (c/8\pi) {\rm Re}(\ee\times \hh^\ast)$ is real. The
Poynting vector ${\bf S}$ involves the non-Hermitian behaviors of
electric and magnetic fields entering it.

As an example, we consider the fields depending on coordinates $y$
and $z$ in an isotropic medium with permittivity $\varepsilon$ in
the form $\hh = H_x(y,z) \eee_x$ and $\ee = E_y(y,z) \eee_y +
E_z(y,z)\eee_z$. The $3\times 3$ stability matrix for the magnetic
field has only two non-zero elements and does not show any peculiar
behavior. The $3\times 3$ matrix $\hat G_E$ for electric field has
zero elements in the first row and in the first column; therefore,
it can be reduced to the $2\times 2$ matrix
\begin{equation}
\hat G_E = \left( \begin{array}{cc} a_{11} & a_{12} \\ a_{21} &
-a_{11},
\end{array} \right)
\end{equation}
where complex-valued elements are equal to $a_{11} = \partial
E_y/\partial y = - \partial E_z/\partial z$, $a_{12} = \partial
E_y/\partial z$, and $a_{21} = \partial E_z/\partial y$. (Here we
exploit the Gauss law for electric field ${\rm div} \ee = \partial
E_y/\partial y + \partial E_z/\partial z = 0$.) This matrix has a
couple of complex eigenvalues $\lambda_{1,2} = \pm \sqrt{a_{11}^2 +
a_{12} a_{21}}$. Exceptional points appear when $\lambda_1 =
\lambda_2 = 0$, that is, $a_{11}^2 + a_{12} a_{21} = 0$. It should
be noticed that the above conditions should hold for non-vanishing
matrix $\hat G_E \neq 0$, because in this case the eigenvectors are
independent.

Richer physics is expected behind the stability matrix for the
Poynting vector $\hat G_S$. For the field we deal with, there are
two components of the energy flux density (we drop the factor
$c/8\pi$ in the definition of the Poynting vector):
\begin{equation}
S_y = {\rm Re}(E_z H_x^\ast), \qquad S_z = - {\rm Re}(E_y H_x^\ast).
\end{equation}
Stability matrix has only four non-zero elements and can be
described by the $2\times 2$ matrix
\begin{equation}
\hat G_S = \left( \begin{array}{cc} a + a_{1} & b + d_1 \\ b - d_2 &
-a + a_{2} \end{array} \right),
\end{equation}
where
\begin{eqnarray}
a &=& -\frac{1}{k_0} {\rm Im}\left( \frac{\partial E_y}{\partial z}
\frac{\partial E_z^\ast}{\partial y} \right), \qquad b = {\rm
Re}\left( \frac{\partial E_z}{\partial z} H_x^\ast
\right), \nonumber \\
a_1 &=& - k_0 {\rm Im}(\varepsilon) |E_z|^2, \qquad a_2 = - k_0 {\rm
Im}(\varepsilon) |E_y|^2, \nonumber \\
d_1 &=& k_0 {\rm Im}(\varepsilon E_y E_z^\ast) \qquad d_2 = k_0 {\rm
Im}(\varepsilon^\ast E_y E_z^\ast).
\end{eqnarray}

Eigenvalues of the matrix $\hat G_S \neq 0$ can be readily
calculated in the general case, but we impose an additional
restriction of lossless medium $\varepsilon = \varepsilon^\ast$.
Then $a_1 = a_2 = 0$ and $d_1 = d_2 = d = k_0 \varepsilon {\rm Im}(
E_y E_z^\ast)$ and the eigenvalues equal
\begin{equation}
\lambda_{1,2} = \pm \sqrt{a^2 + b^2 - d^2}.
\end{equation}
Exceptional point emerges when $d^2 = a^2 + b^2$ and corresponds to
the eigenvalue $\lambda_{EP} = 0$.

Now we consider an example of the field defined as follows
\cite{Novitsky2009}
\begin{eqnarray}
H_x(y,z) &=& A e^{ik z} \left( \kappa - y^2 - \frac{iz}{k}
\right), \nonumber \\
E_y(y,z) &=& \frac{A k}{k_0\varepsilon} e^{ik z} \left( \kappa - y^2
- \frac{ikz+1}{k^2} \right), \nonumber \\
E_z(y,z) &=& \frac{2i A}{k_0\varepsilon} e^{ik z} y,
\label{EMFieldExample}
\end{eqnarray}
where $k = k_0\sqrt{\varepsilon}$ and $A$ is an amplitude of the
magnetic field. Assuming the lossless medium, we calculate the
coefficients for the Poynting vector stability matrix as
\begin{eqnarray}
a = b = \frac{2A^2}{k^2\sqrt{\varepsilon}} z, \quad
d = \frac{2A^2}{k^2\sqrt{\varepsilon}} y \left( k^2 y^2 + 1 - k^2
\kappa \right).
\end{eqnarray}
Coordinates $(y_{EP},z_{EP})$ of the EP position satisfy the
equation
\begin{equation}
|k z_{EP}| = \frac{1}{\sqrt{2}} |k y_{EP} ( k^2 y_{EP}^2 + 1 - k^2
\kappa)|.
\end{equation}

\begin{figure*}[tb]
  \centering
  \includegraphics[width=\linewidth]{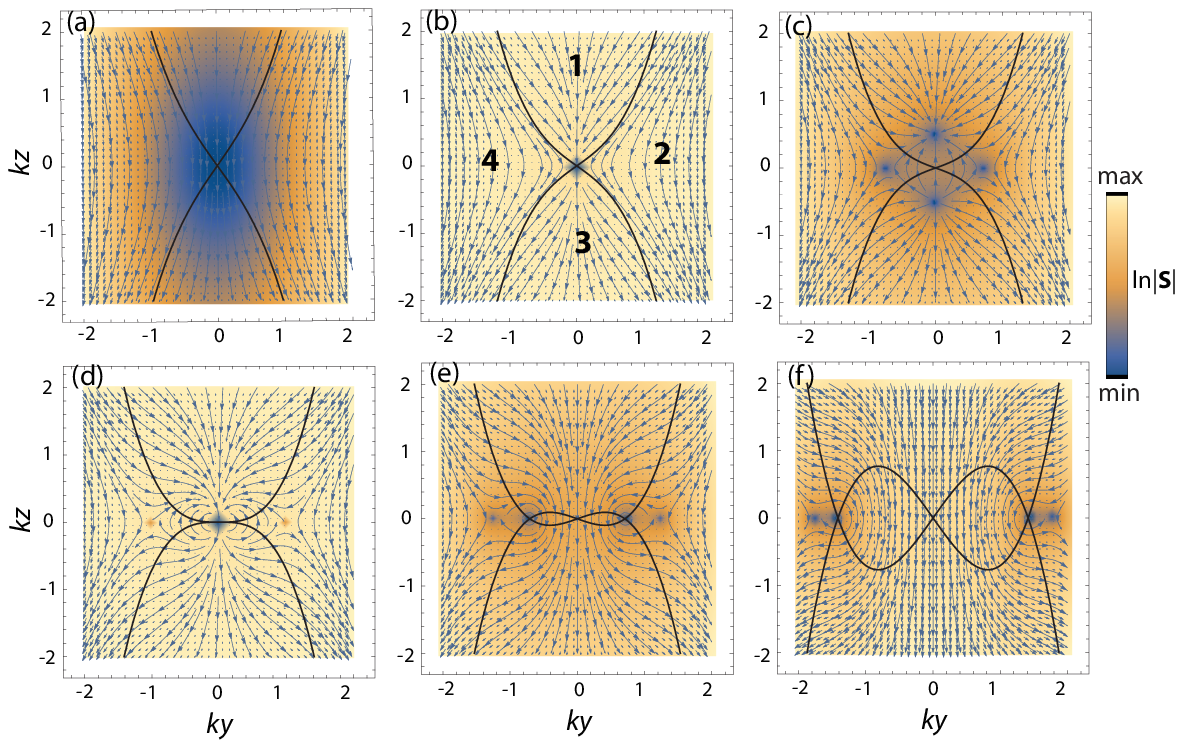}
  \caption{ Map of streamlines of Poynting vector ${\bf S}$ and lines of exceptional
  points (black solid lines) for different parameters $\kappa$: (a) $k^2 \kappa = -1$,
  (b) $k^2 \kappa = 0$, (c) $k^2 \kappa = 0.5$, (d) $k^2 \kappa = 1$,
  (e) $k^2 \kappa = 1.5$, and (f) $k^2 \kappa = 3$. The background shows $\ln|{\bf S}|$ beneath the streamlines.
  }
  \label{fig7}
\end{figure*}

\begin{figure*}[tb]
  \centering
  \includegraphics[width=\linewidth]{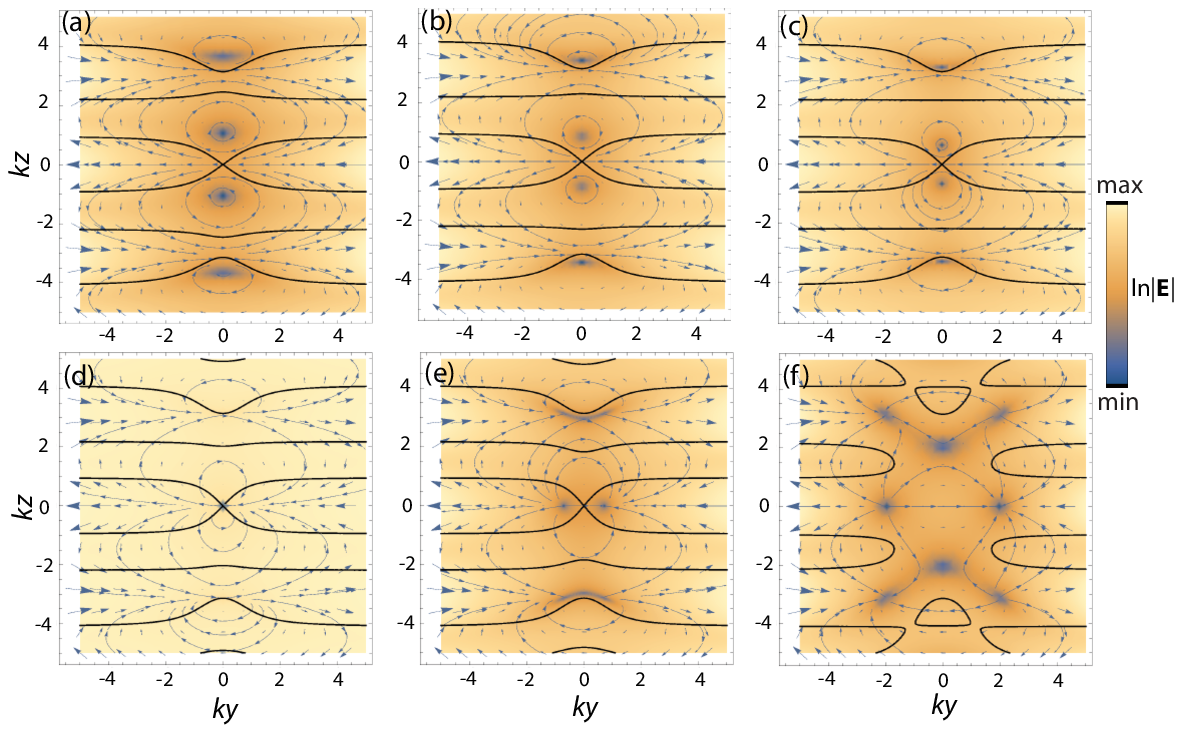}
  \caption{ EP lines (black solid lines) and field lines of real-valued electric fields $\tilde\ee$ for different parameters $\kappa$:
  (a) $k^2 \kappa = -1$,
  (b) $k^2 \kappa = 0$, (c) $k^2 \kappa = 0.5$, (d) $k^2 \kappa = 1$,
  (e) $k^2 \kappa = 1.5$, and (f) $k^2 \kappa = 5$. The background shows the logarithm
  of the electric vector $\ln|{\bf \ee}|$ beneath of the
  streamlines.}
  \label{fig8}
\end{figure*}

The EP lines together with the streamlines of the Poynting vector
and logarithm of the Poynting vector $\ln|{\bf S}|$ are demonstrated
in Fig. \ref{fig7} for various $\kappa$. Parameter $\kappa$ controls
the electromagnetic field and, therefore, the Poynting vector
distribution. When $\kappa$ is negative as in Fig. \ref{fig7}(a),
there are no critical points (centers, focuses, nodes, saddles) in
the Poynting vector distribution and the Poynting vector takes small
non-zero value at the center of the picture. The lines of
exceptional points distinguish four regions marked as 1 to 4 [see
Fig. \ref{fig7}(b)]. Regions on the right 2 and on the left 4
correspond to the Poynting lines which mostly do not cross the EP
lines. The Poynting lines in the other two regions 1 and 3 always
cross the EP lines. Thus, the EP lines split up the ``crossing'' and
''no-crossing`` phases. When $\kappa$ is zero as in Fig.
\ref{fig7}(b), the complex critical point appears in the center of
the panel. The Poynting lines in ``crossing'' phase first converge
in the center and then diverge from it. The phases in this case
clearly confirm their names especially near the center. The complex
critical point splits into four simple critical points conserving
the total Poincar\'{e} index. As it is shown in Fig. \ref{fig7}(c),
two nodes move upwards and downwards, while two saddles shift to the
left and right when $\kappa$ increases. The critical points strongly
perturb the Poynting lines, but their behaviors far from the
critical points stay the same, so that we still can keep the names
``crossing'' and ``no-crossing'' for non-Hermitian phases. The
zeroes of the Poynting vectors are associated with the critical
points. When $k^2 \kappa = 1$ as in Fig. \ref{fig7}(d), the nodes
again merge at the center forming a complex critical point. For
greater $\kappa$, this complex critical point separates into two
centers moving to the left- and right-hand sides. The EP lines
qualitatively change their shape. They still include two upward and
two downward branches that cross each other at the critical points
called centers as it is clearly seen in Fig. \ref{fig7}(e). The EP
lines between the centers do not distinguish different phases with
qualitatively different behaviors of Poynting lines. So, in this
case, we have the ``crossing'' phase in the central region and
``no-crossing'' phase on the left- and right-hand sides. Further
increase of $\kappa$ demonstrated in Fig. \ref{fig7}(f) does not
introduce new features. The critical points move to the left and
right sides, so that the ``crossing'' phase region enlarges.

The electromagnetic field (\ref{EMFieldExample}) can be used for
illustration of the exceptional points in electric field, too. We
will consider the real valued electric field $\tilde\ee = {\rm
Re}\ee$. Then the stability matrix $\hat G_E$ is real-valued as well
with the elements $\tilde a_{11} = {\rm Re}(\partial E_y/\partial
y)$, $\tilde a_{12} = {\rm Re}(\partial E_y/\partial z)$, and
$\tilde a_{21} ={\rm Re}(\partial E_z/\partial y)$. In the case of
the fields (\ref{EMFieldExample}) the condition for the EPs $\tilde
a_{11}^2 + \tilde a_{12} \tilde a_{21} = 0$ reads as
\begin{eqnarray}
2 (ky_{EP})^2 \cos^2(kz_{EP}) = \sin(kz_{EP}) [kz_{EP} \cos(kz_{EP}) \nonumber \\
- (k^2\kappa - (ky_{EP})^2 - 2) \sin(kz_{EP})].
\end{eqnarray}

The EP lines are plotted in Fig. \ref{fig8} as black solid lines.
Owing to the periodicity of the electric fields $\tilde\ee$ the EP
lines repeat. In Fig. \ref{fig8}, we also show the streamlines of
electric fields and the logarithm of the electric field $\ln|\ee|$.
When $\kappa$ is negative as in Fig. \ref{fig8}(a), one observes the
center critical points with the closed loops of streamlines around
them. The EP lines detach the alternating regions with and without
critical points. We can call the non-Hermitian phases as
``localized'' and ``delocalized'' grounding on the behavior of the
streamlines which are respectively circulating around the centers
and passing from the left (right) to the right (left). The similar
behavior is observed in Figs. \ref{fig8}(b) and (c) featuring a
displacement of the singularities toward the EP lines. A
qualitatively different behavior starts from $k^2 \kappa = 1$: All
critical points fall on the EP lines. The centers merge at $z = y =
0$ forming a complex critical point that is split into a couple of
nodes moving to the left and to the right [see Fig. \ref{fig8}(e)].
The other centers after crossing the EP lines transform into the set
of critical points. In general, the critical points change the area
of their location, so that the domains with and without critical
points change places keeping the alternation. Curiously, the
streamlines behaviors in the regions do not change. Now we have
``localized'' phase in the region without critical points and
``delocalized'' phase in the region with critical points. The set of
critical points mentioned above represents nodes and one saddle
point, what is clearly shown in Fig. \ref{fig8}(f). All regions with
critical points join together corresponding to the ``delocalized''
phase. The ``localized'' phases are split forming the regions on the
right- and left-hand sides. Small islands on the upper and lower
parts of the panel do not exhibit any peculiarities and can be
excluded from consideration. Thus, the EP lines for electric field
well distinguish the regions with the critical points and the
regions with distinct behaviors of streamlines. They are completely
different from EP lines of the Poynting vector.

\section{Conclusion}

To sum up, we have presented an overview of approaches to highlight
non-Hermitian behaviors of light fields. First, we have formulated a
non-Hermitian matrix for a plane wave in free space and demonstrated
that it provides an exceptional point separating propagating and
evanescent wave solutions. This approach is tightly related with the
dispersive properties of the waves and can be called
dispersion-based. Second, we have proposed a way to engineer light
beams with imprinted non-Hermiticity. It is realized using a
superposition of plane waves, the vector amplitudes of which are
functions of the non-Hermitian Hamiltonian. The intensity profiles
of such beams exhibit an evident transition between symmetric and
asymmetric field distributions when passing through the exceptional
point. It is revealed that the non-Hermitian features are clearly
observed for non-paraxial light beams, while hidden for paraxial
beams. The properties of light beams with imprinted exceptional
points have been examined for the second- and third-order EPs.
Third, we have developed a theory of non-Hermitian near (local)
electromagnetic fields. We have analyzed the evolution of the maps
of streamlines of the Poynting vector and electric field across a
wide range of parameters and showed how lines of exceptional points
separate the regions of different behavior of streamlines.

We believe that the most promising approach considered here is the
imprinted-non-Hermiticity scheme. If experimentally realized through
phase engineering, it might be exploited for the intricate
light-matter interaction, when the effective non-Hermitian structure
is encoded in the angular spectrum of the field rather than in the
material platform.

%We can also design non-Hermitian light beams in the media.

\section{Acknowledgments}

A.N. thanks the Belarusian Republican Foundation for Fundamental Research (Project No. F26RNF-106), M.R. is grateful to the Russian Science Foundation (Grant No. 25-42-10025), and D.G. acknowledges the National Natural Science Foundation of China (Grant No. 12574342). A.B. acknowledges the National Natural Science Foundation of China (Grant No. W2532010) and the Academic Leadership Program Priority 2030.

%This work was also supported by the National
%Natural Science Foundation of China (Grants No. 12311530763, No.
%12274314, and No. 12474313); Natural Science Foundation of Jiangsu
%Province (Grant No. BK20221240); and a Suzhou Basic Research Project
%(Grant No. SJC2023003).

\appendix

\section{Matrices $A$, $B$, $C$, and $D$ \label{AppA}}

Two-dimensional matrices $A$, $B$, $C$, and $D$ depend on the
parameters of both material and incident plane wave. Material
parameters are defined through the constitutive equations of a
bi-anisotropic medium
\begin{eqnarray}
{\bf D}(\omega, {\bf r}) = \h \E(\omega) {\bf E}(\omega, {\bf r}) + \h \alpha(\omega) {\bf H} (\omega, {\bf r}), \nonumber \\
{\bf B}(\omega, {\bf r}) = \h \kappa(\omega) {\bf E}(\omega, {\bf
r}) + \h \mu(\omega) {\bf H}(\omega, {\bf r}), \label{eq:MaterEq}
\end{eqnarray}
where ${\bf D}$, ${\bf B}$, ${\bf E}$, and ${\bf H}$ are the
electric displacement vector, magnetic induction, electric strength,
and magnetic strength, respectively, $\h \E$ is the dielectric
permittivity tensor, $\h \mu$ is the magnetic permeability tensor,
and $\h \alpha$ and $\h \kappa$ are the gyration pseudotensors.
Incident wave is characterized by the transverse wavenumber ${\bf
k}_t = I {\bf k}$, which is used in dimensionless form as ${\bf b} =
{\bf k}_t/k_0$.

Thereby, the matrices $A$, $B$, $C$, and $D$ take the form
\begin{eqnarray}
\h A&=& {\bf e}_z^\times \h\alpha \h I+ {\bf e}_z^\times
\h\varepsilon{\bf e}_z\otimes\vv_3+({\bf b}+ {\bf e}_z^\times
\h\alpha{\bf e}_z) \otimes\vv_1,\nonumber\\
\h B&=& -{\bf e}_z^\times\h\varepsilon {\bf e}_z^\times+{\bf
e}_z^\times\h\varepsilon{\bf e}_z\otimes {\bf e}_z^\times\vv_4+
({\bf b}+ {\bf e}_z^\times \h\alpha {\bf e}_z)\otimes {\bf e}_z^\times\vv_2,\nonumber\\
\h C&=& \h I \h\mu \h I + \h I \h\mu {\bf e}_z \otimes\vv_1
+(-{\bf a}+ \h I \h\beta{\bf e}_z)\otimes\vv_3,\nonumber\\
\h D&=&-\h I \h\beta {\bf e}_z^\times + \h I \h\mu{\bf e}_z \otimes
{\bf e}_z^\times \vv_2+(-{\bf a} + \h I\h\beta {\bf e}_z)\otimes
{\bf e}_z^\times\vv_4 \label{ABCD_planar}
\end{eqnarray}
where ${\bf e}_z^\times$ is the tensor dual to vector ${\bf e}_z$
\cite{Fedorov1976,Borzdov1997} [$({\bf q}^\times)_{ik} =
\varepsilon_{ijk} q_j$, $\varepsilon_{ijk}$ is the antisymmetric
Levi-Civita tensor and summation over repeated indices from 1 to 3
is assumed], $\bb a = \bb b \times {\bf e}_z$, and
\begin{eqnarray}
\vv_1 &=& \delta_z(\beta_z{\bf e}_z \h\alpha \h I-\varepsilon_z
{\bf e}_z \h\mu \h I-\beta_z {\bf a}),\nonumber \\
\vv_2 &=& \delta_z(\beta_z {\bf e}_z \h\varepsilon \h
I-\varepsilon_q {\bf e}_z \h\beta \h I-\varepsilon_q {\bf a}),\nonumber\\
\vv_3 &=& \delta_z(\alpha_z {\bf e}_z \h\mu \h I-\mu_z {\bf e}_z
\h\alpha \h I+\mu_z {\bf a}),\nonumber \\
\vv_4 &=& \delta_z(\alpha_z {\bf e}_z \h\beta
\h I-\mu_z {\bf e}_z \h\varepsilon I+\alpha_z {\bf a}),  \nonumber \\
\delta_z &=& (\varepsilon_z \mu_z-\alpha_z\beta_z)^{-1}, \quad
\varepsilon_z={\bf e}_z\h\varepsilon{\bf e}_z, \nonumber \\
\mu_z &=& {\bf e}_z\h\mu{\bf e}_z, \quad \alpha_z={\bf
e}_z\h\alpha{\bf e}_z, \quad \beta_z={\bf e}_z\h\beta{\bf e}_z.
\nonumber
\end{eqnarray}

\bibliographystyle{apsrev4-2}
\bibliography{references-v3}

%\bibliography{novitsky}% Produces the bibliography via BibTeX.

\end{document}